\documentclass[aps,pra,onecolumn,superscriptaddress,showkeys,showpacs]{revtex4-2}
\usepackage{graphicx,bm}
\usepackage{float}
\usepackage{subfigure}
\usepackage{amsmath}
\usepackage{amsfonts}
\usepackage{tikz}
\usepackage{relsize}
\usetikzlibrary{shapes.geometric, calc}
\usepackage{inputenc}
\usepackage{amsthm, amssymb,bbm,bm}
\usepackage{hyperref}
\usepackage{siunitx}
\usepackage{tabularx}
\usepackage{multirow}
\usepackage{color}

\newcolumntype{Y}{>{\centering\arraybackslash}X}

\newcommand{\abs}[1]{\left\vert#1\right\vert}

\begin{document}
\title{Single-photon nonlinearities and blockade from a strongly driven photonic molecule}
%
\author{Davide Nigro}
\affiliation{Dipartimento di Fisica, Universit\`{a} di Pavia, via Bassi 6, I-27100 Pavia, Italy}
\author{Marco Clementi}
\affiliation{Dipartimento di Fisica, Universit\`{a} di Pavia, via Bassi 6, I-27100 Pavia, Italy}
\affiliation{Photonic Systems Laboratory (PHOSL), Swiss Federal Institute of Technology Lausanne (EPFL), CH-1015 Lausanne, Switzerland}
\author{Camille-Sophie Brès}
\affiliation{Photonic Systems Laboratory (PHOSL), Swiss Federal Institute of Technology Lausanne (EPFL), CH-1015 Lausanne, Switzerland}
\author{Marco Liscidini}
\affiliation{Dipartimento di Fisica, Universit\`{a} di Pavia, via Bassi 6, I-27100 Pavia, Italy}
\author{Dario Gerace}
\affiliation{Dipartimento di Fisica, Universit\`{a} di Pavia, via Bassi 6, I-27100 Pavia, Italy}

\begin{abstract}
Achieving the regime of single-photon nonlinearities in photonic devices just exploiting the intrinsic high-order susceptibilities of conventional materials would open the door to practical semiconductor-based quantum photonic technologies. Here we show that this regime can be achieved in a triply resonant integrated photonic device made of two coupled ring resonators, without necessarily requiring low volume confinement, in a material platform displaying an intrinsic third-order nonlinearity. By strongly driving one of the three resonances of the system, a weak coherent probe at one of the others results in a strongly suppressed two-photon probability at the output, evidenced by antibunched second-order correlation function at zero-time delay under continuous wave driving.
\end{abstract}
\maketitle

\section{Introduction}
\begin{figure}
    \centering
    \includegraphics[scale=1]{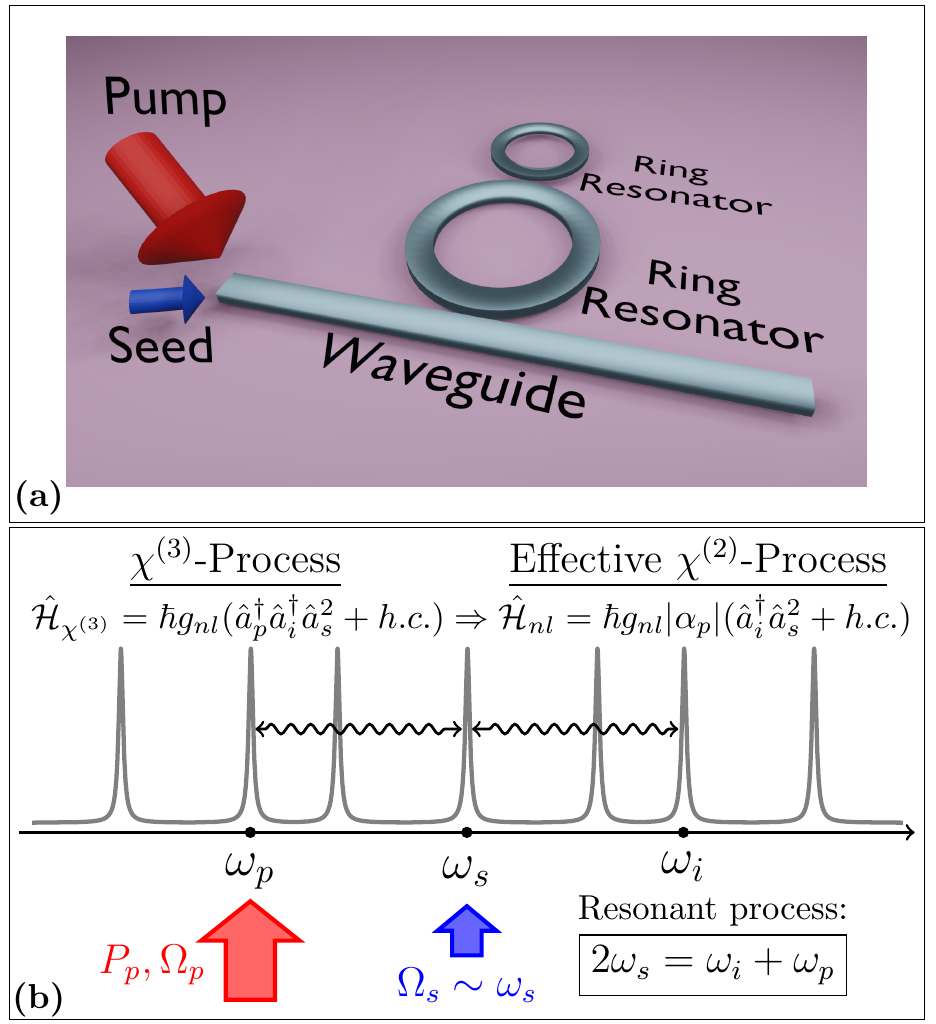}
    \caption{$\textbf{(a)}$ Sketch of the proposed experiment: a single-mode waveguide, driven by a strong and weak coherent field, the Pump and Seed sources respectively, is evanescently coupled to an asymmetric photonic molecule with Kerr-type nonlinearity. $\textbf{(b)}$ Distribution of resonances and transitions (curly arrows) involved in the effective $\chi^{(2)}$ nonlinear process.}
    \label{fig:figure1}
\end{figure}

Reaching the ultimate limit of single-photon nonlinearity in conventional semiconductor materials is the key to a number of applications in quantum technology based on integrated photonic devices.\cite{Wang2020_review} The regime of single-photon nonlinearity would allow to control the response to an external driving, e.g., through the presence of a single photon in a quantum optical device\cite{Chang2007,Gerace2009}. The extreme realization of such a single-photon switch occurs in the so-called ``photon blockade'' (PB) regime, originally proposed for a cavity enhanced optical nonlinearity\cite{Werner1999}: the presence of two radiation quanta within the resonator produces a nonlinear shift that is larger than the cavity mode linewidth, which prevents a second photon from entering the resonator until the first one has been released. The effect was first observed by measuring antibunched second-order correlation of transmitted photons from an optical cavity containing streams of single atoms in strong coupling with the cavity field,\cite{Birnbaum2005} and later in semiconductor cavities coupled to single quantum dots \cite{Faraon2008,Reinhard2012}. 
More recently, signatures of antibunching in the transmitted radiation have also been probed in polariton-confined microcavity samples \cite{Delteil2019,Munoz-Matutano2019}.
All these demonstrations of PB rely on the presence of a strong dipolar interaction of an optically active transition and a tightly confined electromagnetic field. However, it would also be desirable to achieve the regime of strong photon nonlinearities in purely passive materials, only exploiting the intrinsic bulk nonlinear susceptibilities,\cite{Ferretti2012,Majumdar2013} although no significant experimental result has been reported, to date. These approaches typically require engineering low-volume cavities with large quality factors,\cite{Choi2017} such that the resonance shift associated with the effective nonlinear response is of the order of its linewidth \cite{Ferretti2012}. 
We hereby explore a different approach to induce a single-photon nonlinearity on a weak probe beam. The scheme relies on a strong coherent driving of a triply resonant nonlinear system using an intense pump beam, in which enhanced four-wave mixing occurs if the three resonances are equally spaced in frequency, a condition easily met in whispering gallery mode resonators. At difference with previous PB proposals, here we do not require the resonant device to be made of small volume cavities. In fact, a possible experiment is represented in Fig.~\ref{fig:figure1}: an asymmetric photonic molecule made of two ring resonators with different radii results in an inhomogeneous frequency comb with at least three {equally-spaced} resonances that {can be} spectrally isolated from the other nearly resonant modes. A similar device has been proposed for enhanced classical nonlinear processes,\cite{Gentry2014,Heuck2019} or the generation of squeezed light.\cite{Zhang2021}

We show that this device can implement an effective second-order nonlinear response on a weak probe, whose strength depends on the intensity of the pump, and it thus allows one to produce PB on the probe according to the very same mechanism studied before.\cite{Majumdar2013} A similar model has been previously proposed for dipolaritons in a planar microcavity\cite{Kyriienko2014b}. Our results apply instead to passive nonlinear devices, which can be implemented via integrated photonic technologies in material platforms that not only possess a strong Kerr nonlinearity -- such as Lithium Niobate (LiN),\cite{He2019}, Silicon Nitride (SiN) and Silicon-Rich Nitride (SRN),\cite{Ji2017,Wang2015,Bucio2020}, Silicon Carbide (SiC)\cite{Guidry2020}, and III-V semiconductors\cite{Wilson2020,Xie2020,Marty2021} -- but are also operated in their transparency region, with negligible nonlinear absorption. This is crucial to avoid possible heating and saturation of the pump, which would prevent reaching the strong single-photon nonlinear regime. In addition, our theoretical proposal acquires significant experimental relevance in light of the recent demonstrations of strong coupling\cite{Guo2016} and strong nonlinear coupling\cite{Ramelow2019} in ring resonators.

\section{Model and results}
\label{sec:model} 
The device hereby proposed is schematically depicted in Fig.~\ref{fig:figure1}(a). It consists of a photonic molecule realized by two evanescently coupled ring resonators in the ``snowman" configuration, the larger having radius $R$ and the smaller $R/2$. The two rings are evanescently side-coupled to a ridge waveguide, which serves as an input-output channel to either inject excitations in the systems or probe their quantum dynamics via output correlation measurements. In our picture, we assume the ``pump", ``seed", and ``idler" resonance frequencies, respectively $\omega_{p,\,s,\,i}$, to fulfill the four-wave mixing condition $2\omega_s=\omega_p + \omega_i$, and to correspond to those obtained by accounting for any self- and cross-phase modulation correction induced by the presence of a strong continuous-wave (CW) driving field at frequency $\Omega_p$, which is used to induce a significant photon occupation of the resonance at $\omega_p$.
In this respect, the snowman configuration allows for both spectrally isolating these three resonances from the rest of the comb, and to realize, at the same time, a subset of resonances fulfilling four-wave mixing condition, as detailed in appendix \ref{app:A}. In fact, if the system is driven only by the Pump CW source, since the resonances of the coupled rings are not equally spaced, the $\Omega_p$ source will not induce any significant population transfer from the $\omega_p$ resonance to nearby ones. It is then reasonable to assume that in the absence of the Seed driving the $p$-resonance is the only one macroscopically populated, with a steady-state characterized by a mean number of photons $\vert\alpha_p\vert^2$ proportional to the input power $ P_p $. Hence, starting from this stationary condition for the $p$-resonance, the presence of a second weak CW source at $\Omega_s\neq \Omega_p$ will not induce any significant depletion of the $\omega_p$ resonance. Nevertheless, for $\Omega_s\sim \omega_s$, due to the resonant condition, as schematically represented in Fig.~\ref{fig:figure1}(b), such CW source will induce a net population of the seed resonance. We will show that the latter undergoes PB, by characterizing the output statistics at the seed frequency (assuming the pump photons to be fully filtered out).\\
In the rotating frame defined by the two CW source frequencies, for $P_p\gg P_s$, the system Hamiltonian reads (see appendix \ref{app:B} and appendix \ref{app:D} for derivation details):
\begin{equation}\label{eq:Hamiltonian_model}
\hat{\mathcal{H}}=\hbar\Delta( \hat{a}_s^{\dagger}\hat{a}_s + 2\hat{a}_i^{\dagger}\hat{a}_i)+ i\hbar\sqrt{\gamma}(\,F_s\,\hat{a}_s^{\dagger}-F^*_s\,\hat{a}_s)+ \hat{\mathcal{H}}_{nl},
\end{equation}
with $\hat{a}_{\sigma}$ and $\hat{a}_{\sigma}^{\dagger}$ representing annihilation and creation (bosonic) operators at frequency $\omega_{\sigma}$  ($\sigma=s,i$), with $\Delta=\omega_s-\Omega_s$, $F_s$ and $\gamma$ denoting the frequency detuning, the field amplitude of the CW source at $\Omega_s$, and molecule-waveguide input coupling rate (assumed to be equal for the three modes), respectively. Finally, in Eq.~(\ref{eq:Hamiltonian_model}) the nonlinear Hamiltonian reads
\begin{equation}
\hat{\mathcal{H}}_{nl}=\hbar \tilde{g}_{nl}(\hat{a}_s^{\dagger\,2}\hat{a}_i+\hat{a}_i^{\dagger}\hat{a}_{s}^2),
\end{equation}
which describes the effective $\chi^{(2)}$ nonlinear interaction mediated by the $p-$resonance, where $\tilde{g}_{nl}=g_{nl}\vert \alpha_p \vert $ denotes the effective nonlinear coupling, dependent on both the bare $\chi^{(3)}$-nonlinearity $g_{nl}$ and the pump mean-occupation via $\vert \alpha_p \vert$.
To account for optical lossess, we address the convergence towards the stationary condition induced by the CW field at $\Omega_s$ by means of a master equation in the Lindblad form \cite{breuer2002theory}, solving for the unique steady-state solution for the system density matrix (see e.g. appendix \ref{app:C} for a discussion about uniqueness), $\rho_{ss}=\lim_{t\to+\infty}\rho(t)$, where 
\begin{equation}\label{eq:master_equation_general}
\frac{d}{dt}\rho(t)=-\frac{i}{\hbar}\left[\hat{\mathcal{H}},\,\rho(t)\right]
+\sum_{\sigma=s,i} \Gamma_{\sigma}\left[\hat{a}_{\sigma}\rho(t)\hat{a}_{\sigma}^{\dagger}-\frac{1}{2}\left\{\hat{a}_{\sigma}^{\dagger}\hat{a}_{\sigma},\rho(t)\right\}\right]
\end{equation}
with $\Gamma_{\sigma}$ denoting the total linewidth of the $\sigma$-resonance. One can express $\Gamma_{\sigma} = \Gamma^{(0)}_{\sigma}+\gamma$, with $\Gamma^{(0)}_{\sigma}$ being the intrinsic resonance linewidth. In the following, we will assume, for the sake of clarity and without loss of generality, that $\Gamma^{(0)}_{s} = \gamma$ (critical coupling condition), so that $\Gamma_{s}=2\Gamma^{(0)}_{s}$.

The observables of interest are the steady-state average occupation of the seed resonance, $\langle\hat{n}_s\rangle\equiv\langle \hat{a}^{\dagger}_{s}  \hat{a}_{s}\rangle=\mbox{Tr}[\hat{a}^{\dagger}_{s} \hat{a}_{s} \rho_{ss}]$, and the equal time steady-state second-order autocorrelation function, that is $g^{(2)}(0)$, whose expression reads
\begin{equation}
g^{(2)}(0)=\frac{\langle \hat{a}^{\dagger}_{s}\hat{a}^{\dagger}_{s}  \hat{a}_{s}\hat{a}_{s}\rangle}{\langle \hat{a}^{\dagger}_{s}  \hat{a}_{s}\rangle\langle \hat{a}^{\dagger}_{s}  \hat{a}_{s}\rangle}=\frac{\mbox{Tr}[\hat{a}^{\dagger}_{s}\hat{a}^{\dagger}_{s} \hat{a}_{s} \hat{a}_{s} \rho_{ss}]}{\mbox{Tr}[\hat{a}^{\dagger}_{s} \hat{a}_{s} \rho_{ss}]^2}.
\end{equation}

Numerical results obtained for such quantities as a function of the dimensionless nonlinearity,  $\tilde{g}_{nl}/\Gamma_s=g_{nl}\vert \alpha_p \vert/\Gamma_s$, are reported in Fig.~\ref{fig:figure2}. Some further results and comments concerning the numerical simulations are reported in appendix \ref{app:E}\\
In the absence of nonlinearity, since the seed and idler resonances are decoupled, the seed corresponds to the stationary state of a standard driven-dissipative oscillator. This is confirmed by the behavior displayed by both the mean occupation and the autocorrelation function, the former being characterized by a standard Lorentzian lineshape, i.e.,
$\langle\hat{n}_s\rangle={\gamma \vert F_s\vert^2}/\left[{\Delta^2+\left(\Gamma_s/2\right)^2}\right]$ and $g^{(2)}(0)=1$. The latter condition is the fingerprint of the steady-state being in a pure coherent state, whose properties are uniquely determined by the balance between driving and losses. Note that in the present case, due to the weak-pumping of the seed resonance, there is no appreciable resonance splitting, in contrast to what has been reported in the strong-nonlinear regime \cite{Ramelow2019}. The latter is indeed evidenced at stronger seed driving, as shown, e.g., in appendix \ref{app:E}.\\
On increasing the effective nonlinearity, the system approaches a stationary configuration  characterized by a non-trivial degree of correlation, crossing the threshold value $g^{(2)}(0)=0.5$ (i.e., corresponding to the $n=2$ Fock state) for effective nonlinearities $\tilde{g}_{nl}/\Gamma_s > 0.325$. 
We notice that the average photon occupation of the seed mode is only slightly affected by the nonlinearity (see inset), while the different autocorrelation curves display a strong dependence on $\tilde{g}_{nl}/\Gamma_s$. 
Notably, the pronounced dip in the vicinity of $\Delta=0$ is a signature of a strong antibunching regime, induced by the effective $\chi^{(2)}$-nonlinearity. 
In other words, our numerical results show that the effective $\chi^{(2)}$-nonlinearity is responsible for a clear suppression of the occupation of any state containing more than a single seed-photon.

\begin{figure}
\centering
\includegraphics[scale=1]{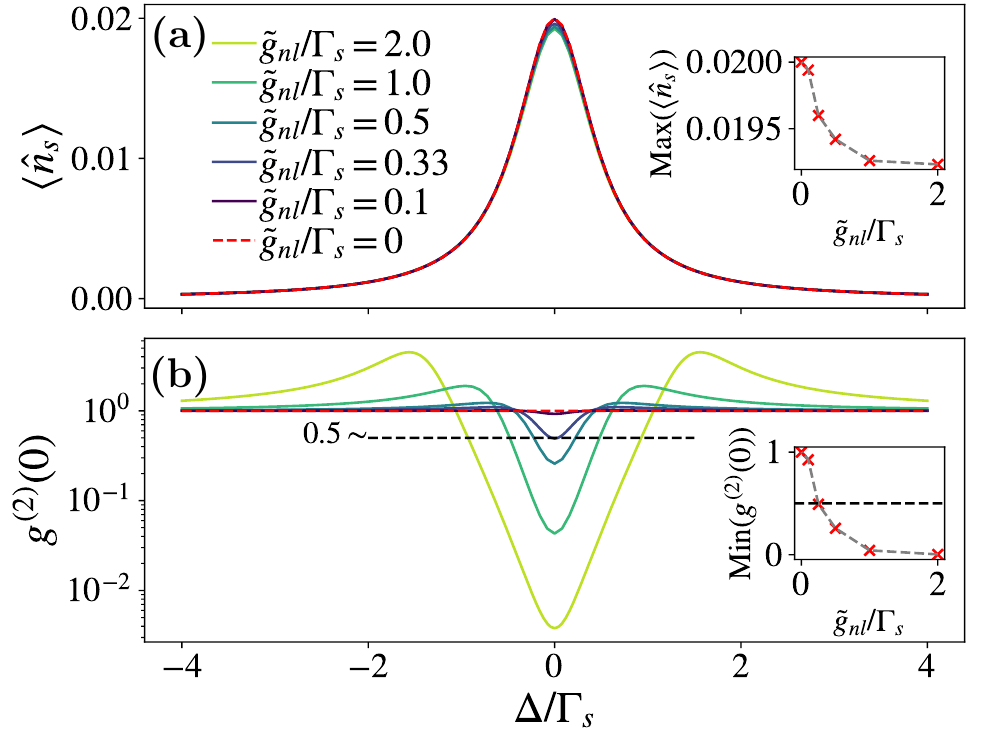}
\caption{$\textbf{(a)}$: Behavior of the steady-state mean occupation $\langle \hat{n}_{s}\rangle$ as a function of $\Delta/\Gamma_s$. The inset shows the behavior of the maximum seed occupation $\mbox{Max}(\langle \hat{n}_{s}\rangle)$ as a function of $\tilde{g}_{nl}/\Gamma$. $\textbf{(b)}$: Steady-state equal time autocorrelation $g^{(2)}(0)$ as a function of $\Delta/\Gamma_s$. The different curves correspond to the different values of $\tilde{g}_{nl}/\Gamma_s$ reported in panel $(\textbf{a})$. The other simulation parameters are: $\Gamma_i=\Gamma_s$, $\gamma/\Gamma_s=0.5$ and $F_s/\sqrt{\Gamma_s}=0.1$.}\label{fig:figure2}
\end{figure}
\begin{table*}[h]
\centering
\small
\begin{tabularx}{\textwidth}{| X |Y Y Y Y|Y| Y| Y |Y|}
\hline
\multirow{2}{*}{Material}  & 
\multirow{2}{*}{$n$} & 
$n_2$  & 
$Q_0$ & 
$R$ &
$\gamma_{nl}$ & 
\multicolumn{3}{c|}{$\tilde{g}_{nl}/\Gamma_s$} \\  
\cline{7-9} 
& 
& 
$\rm{(m^2 W^{-1})}$ & 
$\times 10^6$ & 
$ \rm{(m)}$ &
$\rm{(m^{-1} W^{-1})}$ & 
$P_p=\SI{0.1}{\watt}$ & 
$P_p=\SI{1}{\watt}$ & 
$P_p=\SI{10}{\watt}$ \\ 

\hline

$\mathrm{LiNbO_3}$ \cite{He2019} & 
$2.24$ & 
$1.8 \times 10^{-19}$ & 
$4.4$ & 
$80\times 10^{-6} $ &
$0.7$ & 
$5.7 \times 10^{-4}$ &
$1.8 \times 10^{-3}$ &
$5.7 \times 10^{-3}$ \\

$\mathrm{SRN}$ \cite{Wang2015} & 
$3.1$ & 
$2.8 \times 10^{-17}$ & 
$0.06$ & 
$100 \times 10^{-6}$ &
$550$ & 
$7.3 \times 10^{-4}$ &
$2.3 \times 10^{-3}$ &
$7.3 \times 10^{-3}$ \\

$\mathrm{SiC}$ \cite{Guidry2020} & 
$2.6$ & 
$6.9 \times 10^{-19}$ & 
$1.1$ &
$50\times 10^{-6}$ &
$3.2$ & 
$3.8 \times 10^{-4}$& 
$1.2 \times 10^{-3}$& 
$3.8 \times 10^{-3}$\\

$\mathrm{GaP}$ \cite{Wilson2020} &
$3.3$ &
$1.1 \times 10^{-17}$ &
$0.25$ &
$50\times 10^{-6}$ &
$250$ & 
$4.1 \times 10^{-3}$ & 
$1.3 \times 10^{-2}$ & 
$4.1\times 10^{-2}$ \\

$\mathrm{Si_3N_4}$ \cite{Ji2017} &
$2.0$ & 
$2.5 \times 10^{-19}$ & 
$37$ & 
$115\times 10^{-6}$ &
$0.7$ & 
$1.2 \times 10^{-2}$& 
$3.9 \times 10^{-2}$& 
$0.12$\\

$\mathrm{AlGaAs}$ \cite{Xie2020} & 
$3.3$ & 
$2.6 \times 10^{-17}$ & 
$3.14$ &
$28 \times 10^{-6}$ & 
$390$ & 
$0.23$ & 
$0.73$ &   
$2.3$ \\

\hline

\end{tabularx}
\caption{State of the art parameters and estimated nonlinear coupling for microring resonators 
in different material platforms, comparing linear ($n$) and nonlinear refractive indices ($n_2$), intrinsic quality factor in the resonators ($Q_0$), radius ($R$), and nonlinear parameter ($\gamma_{nl}$). The nonlinear coupling, $\tilde{g}_{nl}$, is estimated for different values of input pump power ($P_p$), and assuming a working frequency $\omega_s=\SI{193}{\tera\hertz}$, with $\Gamma_s=\omega_s/Q_0$.
The values of group velocity ($v_g$) and effective area ($A_{eff}$) required to estimate $\gamma_{nl}$ were numerically calculated via eigenmode expansion using Ansys Lumerical software.}
\label{tab:comparison_gamma}
\end{table*}
\section{Discussion}
\label{sec:discussion} 
We now consider the accessibility of the experiment described above by integrated photonic technologies at the state of the art and in different material platforms of current interest.
First, we directly compare different platforms typically used in nonlinear integrated photonics in Table \ref{tab:comparison_gamma}. The choice has been made based on those materials presenting significant nonlinear response, low propagation losses, and absence of two-photon absorption in the typical telecom band. This explains the exclusion of Silicon, for example. Based on these figures of merit, the same platforms are commonly used for the generation of optical frequency combs.
From the reported values of the (intrinsic) quality factor $Q_0=\omega_s/\Gamma^{(0)}_{s}$, we estimate the coupling rate according to the following expression \cite{Ramelow2019} $g_{nl}/\Gamma_s=\hbar\omega_s v_g^2 \gamma_{nl} /(2\pi R \Gamma_s)$, where $v_g$ is the group velocity and $\gamma_{nl}=\frac{\omega_s n_2}{A_{eff} c} $ is the nonlinear parameter of the waveguide related to its $\chi^{(3)}$ response, which accounts for the enhancement effect due to spatial confinement, where $n_2$ is the nonlinear refractive index, and the effective mode area calculated for the given ring resonator is defined as $A_{eff}=\left(\int \abs{E}^2 dS\right)^2/\int_{ring} \abs{E}^4 dS$, in which the steady state electric field profile is integrated on a surface domain encompassing the ring cross-section (the denominator integral is only performed on the nonlinear material domain, i.e. the ring cross section).
From the above expressions, it is possible to estimate the nonlinear coupling rate as:
\begin{equation}\label{eq:explicit_nonlinearity_VS_threshold_OPO}
    \frac{\tilde g_{nl}}{\Gamma_s} = 
    \frac{g_{nl}}{\Gamma_s} \sqrt{\frac{P_p}{\hbar\omega_s\Gamma_s}}    =
    \frac{\gamma_{nl}v_g^2}{2\pi R}
    \sqrt{\frac{\hbar\omega_s P_p}{\Gamma_s^3}}
\end{equation}

\begin{figure}
    \centering
    \includegraphics[scale=0.8]{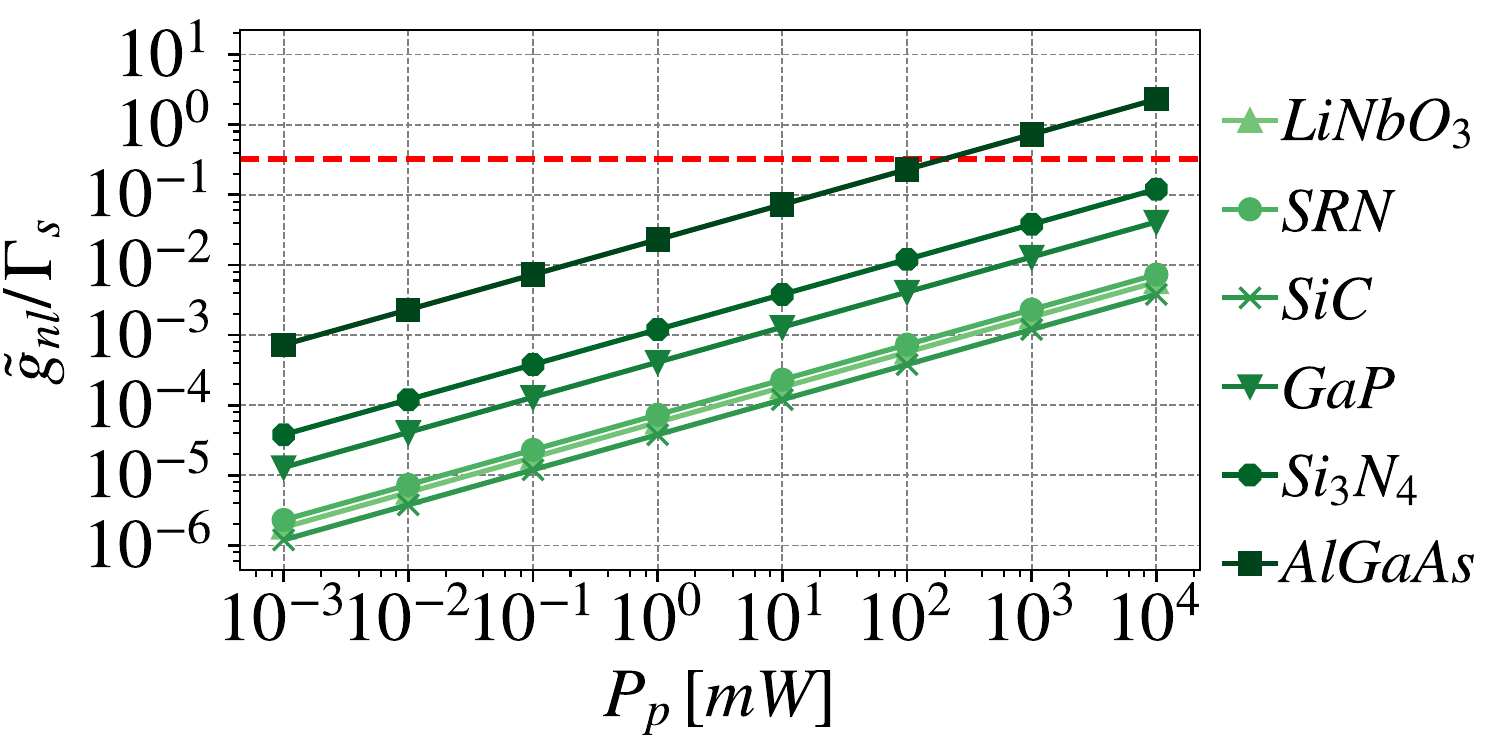}
    \caption{Behavior of the effective nonlinearity $\tilde g_{nl}/\Gamma_s$ (logscale) as a function of the pump power $P_p\,[mW]$ for the material platforms reported in Table \ref{tab:comparison_gamma}. The dashed horizontal line locates the threshold value $\tilde{g}_{nl}/\Gamma_s=0.325$, leading to a $g^{(2)}(0)=0.5$.}
    \label{fig:figure3}
\end{figure}

We report in Table \ref{tab:comparison_gamma} realistic values for the relevant parameters appearing in Eq.~\ref{eq:explicit_nonlinearity_VS_threshold_OPO}, as well as a few estimates for the effective nonlinerarity at different pump powers ($P_p=0.1 W,\,1W,\,10W$), for different material platforms. The expected behavior for the effective nonlinearity in the same platforms as a function of driving powers in the  $1\mu W$ - $10\,W$ range is shown in Fig.~\ref{fig:figure3}. 
First of all, we note that all of the values listed in the Table \ref{tab:comparison_gamma} are within 3 orders of magnitude from the ``nonclassical limit" $\tilde{g}_{nl}/\Gamma_s=0.325$ (threshold value corresponding to the dashed horizontal line in Fig.~\ref{fig:figure3}), whereas the highest calculated values largely exceed the best result achieved in high-$Q$, $\chi^{(2)}$ resonators\cite{Lu2020} even at relatively low pump power.
The threshold for single-photon antibunching at $g^{(2)}(0)=0.5$ is clearly reached and overcome, for the input pump powers considered here, in the AlGaAs platform \cite{Xie2020}, which successfully combines a high nonlinearity with a tight temporal and spatial confinement. Further improvements on the other material platforms are expected by reducing the microring radii while keeping large quality factors.
These results suggest that single photon blockade in passive $\chi^{(3)}$ devices might realistically be achieved in current fabrication technology, upon appropriate engineering of the nonlinear device. More in general, a high figure of merit is attained in III-V alloys \cite{Wilson2020,Marty2021}, for which bandgap engineering allows to maximize the nonlinear index while avoiding two-photon absorption.
In this perspective, the recently developed frequency combs in bichromatic photonic crystal resonators \cite{Alpeggiani2015,Clementi2020,Marty2021} represent a promising approach to obtain a triplet of equally-spaced resonances, as compared to the ring resonator layout described in this work, where the intrinsically low mode volume would contribute to increase the nonlinear coupling. For completeness, an estimate of the nonlinear parameters for such microresonators has been numerically obtained from data reported in \cite{Marty2021}, resulting in, e.g., in $\tilde{g}_{nl}/\Gamma_s\simeq 0.02$ (0.064) at $P_p=0.1$ W (1 W). 
Furthermore, a similar approach could also be investigated for intrinsic $\chi^{(2)}$ processes in doubly-resonant photonic crystal resonators \cite{Wang2020}.
The SiC \cite{Guidry2020} and SiN \cite{Ji2017} platforms have recently gained popularity in integrated photonics, thanks to their excellent optical and nonlinear properties, and to CMOS compatibility. They also represent two promising candidates for quantum technology applications relying on photon blockade, although significant improvements in respectively $Q$ factor and nonlinearity would be needed to reach a performance comparable to the one of III-V devices.
From a technological perspective, the performance of the SiN platform could be dramatically improved by using non-stoichiometric depositions (SRN)\cite{Bucio2020}, for which improvements in the $n_2$ index exceeding two orders of magnitude have been demonstrated \cite{Wang2015}, without incurring in nonlinear absorption.
Last but not least, the LiN thin-film platform \cite{He2019} represents a promising candidate for  applications in integrated quantum- and nonlinear-photonics. While in this case the nonlinearity is quite small as compared to III-V semiconductors, the large second-order nonlinear response of this material might be exploited to achieve more complex interactions involving both the second- and third-order susceptibilities.
As a final comment, let us point out that the pump power used to calculate the nonlinear coupling might differ in practical devices, and could be further increased in some of these materials before incurring in the damage threshold. Furthermore, we note that the present treatment, done for a CW pump, could be extended to the pulsed regime, thus allowing to reach higher peak power and therefore increasing the effective nonlinearity achievable.\\

\bibliography{bibliography}

\appendix

\section{Resonance spectrum of the photonic molecule}\label{app:A}
\begin{figure}[h]
\centering
\includegraphics[scale=0.12]{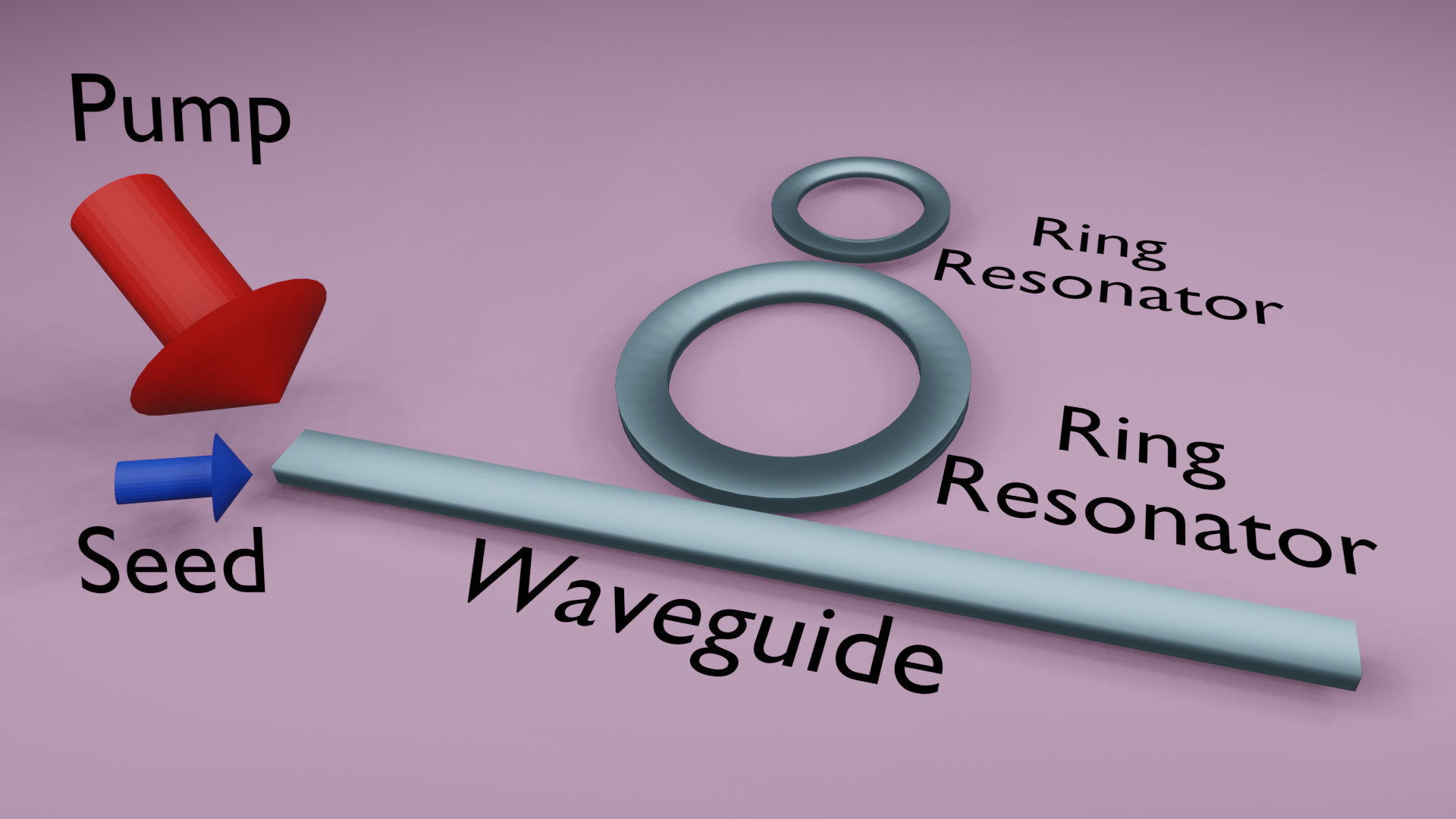}
 \caption{Proposed experimental setup to realize photon blockade in a photonic integrated chip: a single-mode waveguide is evanescently coupled to an asymmetric photonic molecule (coupled ring resonators) with Kerr-type nonlinearity. }\label{fig:exp_setup}
 \end{figure}

\begin{figure}[h]
\centering
\includegraphics[scale=0.95]{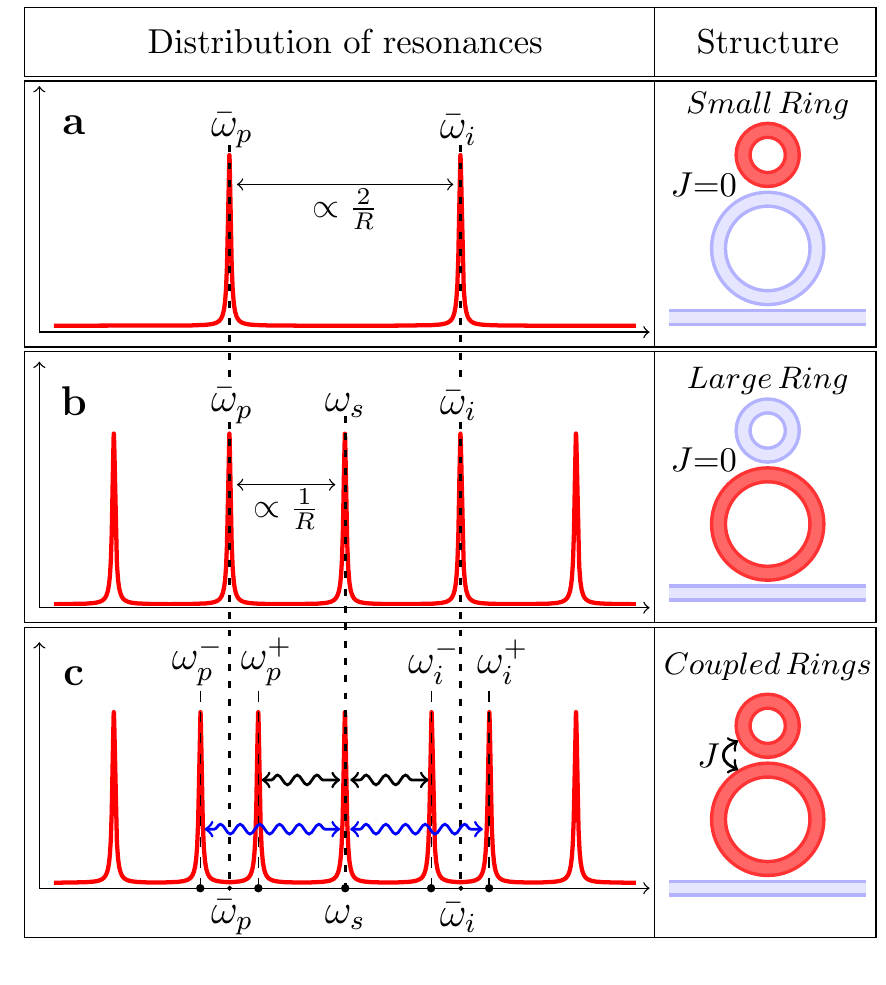}
\caption{Distribution of the resonances of (\textbf{a}) the uncoupled small ring (radius $R/2$), (\textbf{b}) the uncoupled large ring (radius $R$) and of (\textbf{c}) the photonic molecule, with $J$ denoting the coupling strength between the rings. For $J=0$, see panel (\textbf{a}) and panel (\textbf{b}), the resonances in each ring are evenly spaced in frequency and due to the ratio between the radii those at $\bar{\omega}_p$ and $\bar{\omega}_i$ are aligned. For $J\neq 0$, such aligned resonances give rise to the doublets $(\omega^{-}_{p},\omega^{+}_p)$ and $(\omega^{-}_{i},\omega^{+}_i)$. }
\label{fig:ring_resonances}
\end{figure}

In this section, we discuss the motivation for using a device consisting of two coupled ring resonators, such as the photonic molecule considered in the main text, rather than a simpler one consisting of a single ring resonator coupled to a waveguide, in light of engineering strong nonlinear interactions between two given resonances of the structure, mediated by the presence of a macroscopic population of a third resonance.\\
The starting point of our discussion concerns the distribution of resonances in a ring resonator. In an ideal circular (i.e. rotationally invariant) structure, the frequency distribution of azimuthal modes is uniquely determined by $(i)$ the effective index $n_{\rm eff}$ of the waveguide, and $(ii)$ the geometry of the resonator, that is, its radius $R$. By requiring the optical path difference traveled by light along one round-trip of the resonator, that is $n_{\rm eff} 2\pi R$, to be a multiple of the wavelength, and the effect of dispersion, it is straightforward to obtain that resonances are distributed according to:
\begin{equation}\label{eq:energy_spacing}
\omega_m=m \delta \omega= m \frac{c}{n_{\rm eff} R},
\end{equation}
with the azimuthal number $m$ being an integer used to label the set of resonances $\{\omega_m\}$, and with $\delta \omega$ denoting the Free Spectral Range (FSR) of the ring. In such a scenario, due to the equal spacing between consecutive resonances, the maximum population that can be stored within a single resonance is severely limited by four-wave mixing processes converting two photons at the frequency of interest, let us say $\omega=\omega_p$, into a pair of photons corresponding to the adjacent resonances, that is $(\omega_p+ \delta \omega,\,\omega_p- \delta \omega)$. \\ 
In the case of rings with non-commensurate FSRs, a similar limitation would occur also in the device reported in  Fig. \ref{fig:exp_setup}. On the contrary, if the radii of the two resonators are commensurate, as sketched in panel $\textbf{(a)}$ and panel $\textbf{(b)}$ of Fig. \ref{fig:ring_resonances}, some of the resonances (namely those at $\bar{\omega}_p$ and $\bar{\omega}_i$) may share the same resonance frequency, and, due to the interaction between the rings, get hybridized, generating new doublets of resonances. With reference to panel \textbf{c} of Fig.~\ref{fig:ring_resonances}, in the case of perfectly aligned resonances, the position of such doublets is given by:
\begin{equation}\label{eq:hybridized_resonances}
    \omega^{\pm}_{\sigma}=\bar{\omega}_{\sigma}\pm J,\,\,\sigma=p,\,i
\end{equation}
with $J$ denoting the coupling between the rings (which is assumed to be of the same intensity for the resonances at $\bar{\omega}_p$ and $\bar{\omega}_i$). In particular, as explicitly shown in Eq. \ref{eq:hybridized_resonances}, the hybridization of such resonances leads to $(i)$ doublets which are less sensitive to four-wave mixing processes (since adjacent resonances are not evenly spaced) and $(ii)$ to the natural formation of isolated sets of three equally spaced resonances from the comb of each single ring resonator. Indeed, it is easy to verify that (see e.g. panel $(\textbf{c})$ of Fig. \ref{fig:ring_resonances}):
\begin{equation}
    \omega^{-}_{p}+ \omega^{+}_i=2 \omega_s\,\quad \mbox{and}\quad\omega^{+}_{p}+ \omega^{-}_i=2 \omega_s.
\end{equation}
Although in the main text we set $\omega_p=\omega^{-}_{p}$ and $\omega_i=\omega^{+}_i$, the pair of resonances $(\omega_p,\,\omega_i)$ that can be used to generate an effective strong $\chi^{(2)}$ interaction may also correspond to  $(\omega^{+}_p,\,\omega^{-}_i)$. In particular, the analysis shown in the main text would remain the same. In addition, we stress that for the purposes of discussing effective $\chi^{(2)}$ nonlinearities induced by a strong continuous-wave (CW) source driving one of the resonances in the $p$-doublet, one can safely neglect nonlinear effects involving the other resonance. For the sake of clarity, in order to show that we can arbitrarily choose one of the two $p$-resonances in the vicinity of $\bar{\omega}_p$, let us make the opposite choice with respect to the manuscript and assume $\omega_p=\omega^{+}_p$ and $\omega_i=\omega^{-}_i$. In the absence of any other CW source, since the resonances of the coupled rings are not equally spaced, the $\Omega_p$ source acting in the proximity of $\omega_p$ will not induce any significant population transfer from this resonance to nearby ones. Therefore, it is reasonable to assume that in the stationary regime induced by the $\Omega_p$ source such a resonance is the only one macroscopically populated, with a state characterized by a mean amplitude $\alpha_p$, such that $\vert \alpha_p \vert^2\propto P_p $, the latter being the optical power of the strong CW source in the input waveguide. Once such a stationary regime is reached, the presence of a second weak CW source at $\Omega_s\neq \Omega_p$ will not induce any significant depletion of the $\omega_p$ resonance. Nevertheless, for $\Omega_s\sim \omega_s$, such CW source would induce a net population of the seed resonance. In parallel, the nonlinear processes corresponding to the curly arrows in panel $\textbf{c}$ of Fig.~\ref{fig:ring_resonances}, would scatter photons out of the seed resonance. Notice, however, that due to the presence of the strong driving on the pump resonance at $\omega^{+}_p$, nonlinear effects connecting states in the $(\omega^{+}_p,\,\omega_s,\,\omega^{-}_i)$ triplet will be more relevant than those connecting states in the $(\omega^{-}_p,\,\omega_s,\,\omega^{+}_i)$ triplet. As a consequence, these latter contributions, which are controlled by the intrinsic $\chi^{(3)}$ nonlinearity of the material and convert two seed photons into a $(\omega^{-}_{p},\, \omega^{+}_{i})$ pair, can be neglected while describing the dynamics induced by the weak CW source at $\Omega_s$. 
One comes to same conclusions by considering $\omega_p=\omega^{-}_{p}$ and $\omega_i=\omega^{+}_i$ as in the main text.\\
 
\section{Dynamical model of the photonic molecule}\label{app:B}
In the absence of any source term and any coupling to the surrounding environment, that is $F_s=F_p=0$ and $\Gamma_{\sigma}=0 (\sigma=p,\,s,\,i)$, we assume the three resonances of the photonic molecule to be governed by the following Hamiltonian operator 
\begin{equation}\label{eq:Hamiltonian_model}
\hat{\mathcal{H}}=\sum_{\sigma=p,\,s,\,i}\hbar\tilde{\omega}_{\sigma} \hat{a}_{\sigma}^{\dagger}\hat{a}_{\sigma} + \sum_{k,\,l,\,m,\,n=p,\,s,\,i} \left(\hbar g_{klmn}\hat{a}_{k}^{\dagger}\hat{a}_{l}^{\dagger}\hat{a}_{m}\hat{a}_{n}+h.c.\right).
\end{equation}
In our scheme, $\tilde{\omega}_{\sigma}$ denotes the energy of the $\sigma-$resonance of the cold system, that is in the absence of driving terms. The $\omega_{\sigma}$ reported in the main text describe instead the loaded-system resonances. The coupling constant $g_{klmn}$ accounts for nonlinear effects that convert photon pairs occupying the $m$ and $n$ modes to photon pairs in the modes $k$ and $l$, and it is proportinal to the component of the susceptibility tensor $\chi^{(3)}_{klmn}$ of the material used to made the ring resonators. The operators $\hat{a}_{\sigma}^{\dagger}$ and $\hat{a}_{\sigma}$ correspond to creation and annihilation bosonic operators of the $\sigma-$resonance, which satisfy the canonical commutation rules:
\begin{equation}
[\hat{a}_{\alpha},\,\hat{a}_{\beta}]=0,\quad [\hat{a}_{\alpha},\,\hat{a}^{\dagger}_{\beta}]=\delta_{\alpha\,\beta},\quad\,\alpha,\beta=p,\,s,\,i
\end{equation}
with $\delta_{\alpha\,\beta}$ denoting the Kronecker delta, i.e.  $\delta_{\alpha\,\beta}=1\,$ if and only if $\alpha=\beta$.\\
In this framework, once the continuous-wave sources are turned on, some extra terms must be included into the model, and the total Hamiltonian, which becomes explicitly time-dependent, reads:\
\begin{equation}
\hat{\mathcal{H}}_{tot}=\hat{\mathcal{H}}+\hat{\mathcal{H}}_{F},
\end{equation}
with
\begin{equation}\label{eq:driving_terms}
\hat{\mathcal{H}}_{F}=i\hbar\sqrt{\gamma}(\,F_p\,\hat{a}_{p}^{\dagger}e^{-i\,\Omega_p\,t}-F^*_p\,\hat{a}_p e^{i\,\Omega_p\,t})+ i\hbar\sqrt{\gamma}(\,F_s\,\hat{a}_{s}^{\dagger}e^{-i\,\Omega_s\,t}-F^*_s\,\hat{a}_s e^{i\,\Omega_s\,t}),
\end{equation}
where the first term describes the driving of the $p$-resonance by a monochromatic source at frequency $\Omega_p$, and the latter accounts for the driving of the seed resonance with a monochromatic source at frequency $\Omega_s$. In particular, for the sake of clarity, it is worth noting that since $\gamma$ has the dimensions of a rate, that is $\left[ \gamma\right]=s^{-1}$, and since $[\sqrt{\gamma}F_p]=[\omega_{\sigma}]=s^{-1}$, the field amplitudes $F_{\sigma}$ do have the dimensions of the square root of a rate, i.e. $\left[F_{\sigma}\right]=s^{-1/2}$. \\
In a realistic setup, photons can escape from the resonators. As a consequence, the Hamiltonian model in Eq.~\ref{eq:Hamiltonian_model} is not sufficient to describe the time evolution of the photonic system, and one needs to introduce losses in the theoretical description. A possible strategy is to use Lindblad formalism, that is to cast the time evolution of the system density operator $\rho(t)$ in the following form
\begin{equation}\label{eq:master_equation_general}
\frac{d}{dt}\rho(t)=\mathcal{L}[\rho(t)]\equiv-i\left[\hat{\mathcal{H}}_{tot}/\hbar,\,\rho(t)\right]+\sum_{\sigma=p,\,s,\,i}\Gamma_{\sigma}\,\mathbb{D}[\hat{a}_{\sigma};\,\rho(t)],
\end{equation}
with $\mathcal{L}$ being the Lindbladian (super-)operator, and with
\begin{equation}\label{eq:dissipator_expression}
\mathbb{D}[\hat{L};\,\rho(t)]\equiv \hat{L}\rho(t)\hat{L}^{\dagger}-\frac{1}{2}\left\{\hat{L}^{\dagger}\hat{L},\,\rho(t)\right\}
\end{equation}
denoting the dissipator associated to a decay channel governed by the Lindblad operator $\hat{L}$. In particular, in the present case we assume two-photon losses to be negligible and we only consider non-unitary processes associated to one-photon losses, that is those induced by a linear coupling between the resonances and the surrounding environment. Such effects are captured by Lindblad operators proportional to annihilation operators of the molecule modes. In particular, in our model the parameters $\Gamma_{\sigma}=\Gamma^{(0)}_{\sigma}+\gamma$ account for the total decay rate of the resonances, being $\Gamma^{(0)}_{\sigma}$ the resonance linewidth in the absence of coupling with the waveguide.\\

\section{Steady-state configuration and driving protocols}\label{app:C}
\begin{figure}
\centering
\subfigure[$ protocol\,A$]{\includegraphics[scale=0.99]{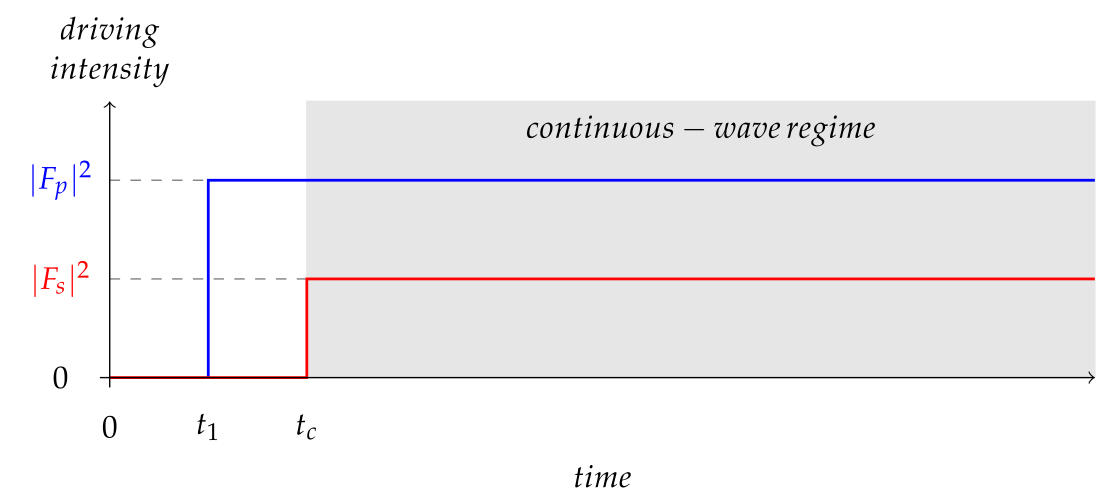}\label{fig:protocol_A}}\\
\subfigure[$protocol\,B$]{\includegraphics[scale=0.99]{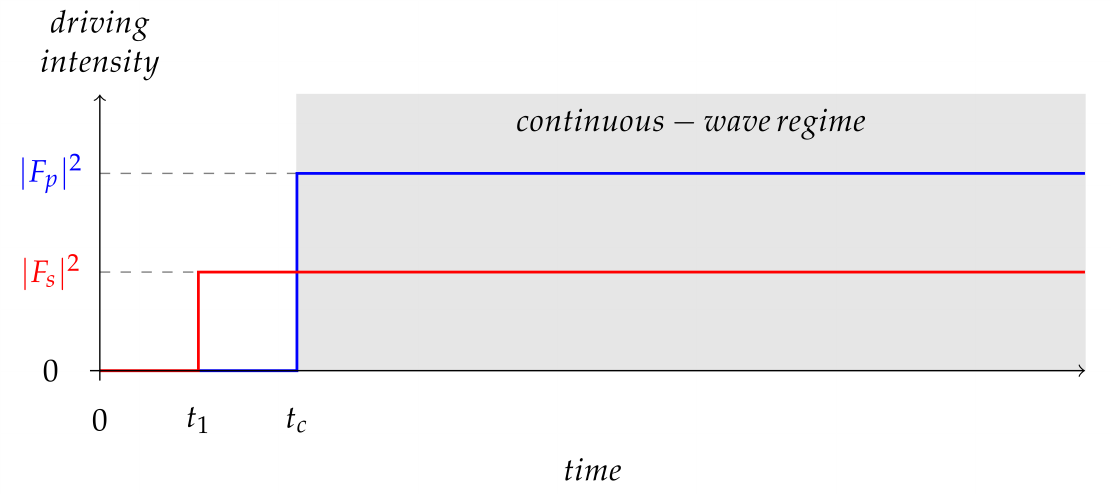}\label{fig:protocol_B}}
\caption{Different driving protocols leading to the same continuous-wave regime for $t>t_c$. Fig.~\ref{fig:protocol_A}: the pump field (with intensity $\propto \vert F_p \vert^2 $) is turned on at $t_1$ and then at later time $t_c$ we start to drive also the population of the $s$-resonance (with intensity $\propto \vert F_s \vert^2 $). Fig.~\ref{fig:protocol_B}: we first turn on the driving of the $s$-resonance  at $t_1$ (with intensity $\propto \vert F_s \vert^2 $) and then at later time we start to drive also the population of the $p$-resonance (with a field intensity $\propto \vert F_p \vert^2 $).}\label{fig:driving_protocols}
\end{figure}

In the continuous-wave regime, the probability distributions of interest are those corresponding to the steady-state configurations. Such configurations, that can be formally defined as follows
\begin{equation}\label{eq:convergence_to_ss}
\lim_{t\to+\infty} \rho(t)=\rho^{(n)}_{ss}\quad n=1,\cdots,\,N_{ss},
\end{equation}
and that correspond to the $N_{ss}$ configurations annihilated by the Lindbladian, that is
\begin{equation}
\frac{d}{dt}\rho^{(n)}_{ss}=0=\mathcal{L}[\rho^{(n)}_{ss}],
\end{equation}
are those states approached by the system when there is a perfect balance between unitary processes and non-unitary ones accounted by the dissipator. In particular, once such configurations are determined, one has access to expectation values by simply computing the following quantities:
\begin{equation}
\langle \hat{O}\rangle^{(n)}=\mbox{Tr}[\hat{O}\rho^{(n)}_{ss}],
\end{equation}
with $\hat{O}$ being an operator and with ``$\mbox{Tr}[A]$" denoting the trace operation over the quantity $A$.\\
As specified discussed in the main text, our final goal is to characterize the stationary regime reached in the presence of a strong driving on the $p$-resonance and a weak one acting on the seed. In terms of the Hamiltonian parameters, such regime corresponds to $\vert F_p\vert^2 \gg \vert F_s\vert^2$.\\
In general, one might legitimately wonder if such a regime defines a unique equilibrium distribution. In other words, one might wonder if different driving protocols corresponding to the same continuous-wave regime, that is equivalent after some reference time $t=t_c$, may lead to different steady-states. In order to clarify this point, please consider Fig.~\ref{fig:driving_protocols}, where we sketched two different protocols that consider different driving combinations for $t<t_c$, but defining the same continuous-wave regime after for larger times. In both cases, let us assume the system to be in the vacuum state at $t=0$. In the ``protocol A", see Fig. \ref{fig:protocol_A}, we first turn on the pump-driving field at $t=t_1$ and then at later time we also drive the population of the $s$-resonance. When using the ``protocol B" we do consider the opposite scenario, so that we first start to drive the seed and after $t=t_c$ we also inject photons in the $p$-resonance. As a result, since the protocols are different, it would not be surprising to have different configurations at $t=t_c$. Nevertheless, what should we expect for $t\gg t_c$?
The answer to this question is that the faith of the (open) quantum system depends on the number of steady-states supported by the master equation for $t> t_c$. In general, if different steady-states exist, then the future evolution of the quantum system is defined by its initial configuration. Indeed, similarly to what happens in the case of closed quantum systems, the existence of many stationary regimes is related to the existence of at least one (strong) $symmetry$ preserved by the Lindbladian. Let us denote such a conserved quantity by $\hat{Q}$. If the two states generated by the two driving protocols, let us say $\rho_{A}(t_c)$ and $\rho_{B}(t_c)$, are characterised by two different values of such symmetry, that is $q_A=\mbox{Tr}[\hat{Q}\,\rho_{A}(t_c)]$ and $q_B=\mbox{Tr}[\hat{Q}\,\rho_{B}(t_c)]$, then due to the conservation of $\hat{Q}$ one can conclude that 
\begin{equation}
\rho^{(A)}_{ss}\neq \rho^{(B)}_{ss}.
\end{equation} 
On the other hand, if the steady-state is unique, the dynamics is \emph{relaxing}, that is
\begin{equation}
\lim_{t\to +\infty}\rho(t) = \rho_{ss}\quad \forall\,\rho_0=\rho(0),
\end{equation}
or in other words, if the steady-state is unique, then any initial configuration will converge to it if we wait a sufficiently long time. Therefore,  for our purposes, it becomes of primary interest to understand if the master equation in Eq. \ref{eq:master_equation_general} does support a unique steady-state in the continuous-wave regime for $t>t_c$. In order to verify that this is indeed the case, it is sufficient to consider the results reported in Ref. \cite{Nigro_2019}, where it is shown that the presence of one-photon losses acting on each resonance of a quantum system is a sufficient condition to conclude that the master equation has a unique steady-state $\rho_{ss}$. Furthermore, by means of the results derived in Ref. \cite{SPOHN1976189}, one can prove that the dynamics encoded into Eq. \ref{eq:master_equation_general} is also \emph{relaxing}. \\
In other words, these facts ensure that:
\begin{itemize}
\item[1)] the stationary state for $t>t_c$, that is the steady-state $\rho_{ss}$, is \emph{unique};
\item[2)] any initial configuration at $t=t_c$ will converge to $\rho_{ss}$ for sufficiently long times;
\item[3)] the dynamics for $t<t_c$ is irrelevant for determining  the long time behavior of the system, in the sense that any deviation introduced by a different choice in the driving protocol for $t<t_c$, for instance ``protocol A" versus ``protocol B", will eventually disappear for large enough time.
\end{itemize}

\section{Derivation of the effective two-resonance model}\label{app:D}
In this section we derive an effective model describing the dynamics of the seed-idler subsystem within the full dynamical regime defined in the previous Section. As a first step, we look for the physical configuration that guarantees the dynamics of such a subsystem to be described by the master equation reported in the main text. To this purpose, let us consider the generic rotating frame defined by the following unitary operator:
\begin{equation}
R(t)=e^{ i(\tilde{\Omega}_{p}\hat{a}^{\dagger}_{p}\hat{a}_p +\tilde{\Omega}_s \hat{a}^{\dagger}_{s}\hat{a}_s +\tilde{\Omega}_i \hat{a}^{\dagger}_{i}\hat{a}_i)t}\equiv\exp(i\hat{\Lambda}t)
\end{equation}
In this frame, the rotated density operator is given by 
\begin{equation}
\tilde{\rho}=R(t)\rho R^{\dagger}(t),
\end{equation}
and its time evolution is given by the following rotated master equation
\begin{equation}
\frac{d}{dt}\tilde{\rho}=i\,R(t)[ \hat{\Lambda},\, \rho]R^{\dagger}(t)+R(t)\frac{d}{dt}\rho(t)R^{\dagger}(t)=iR(t)[ \hat{\Lambda},\, \rho]R^{\dagger}(t) + R(t)\mathcal{L}[\rho(t)]R^{\dagger}(t).
\end{equation}
In order to proceed further, we need an explicit expression for the rotated bosonic operators, that is $\tilde{a}_{\sigma}(t)=R(t)a_{\sigma}R^{\dagger}(t)$. By considering that $\tilde{a}_{\sigma}(t)$ is characterized by the following time evolution
\begin{equation}\label{eq:rotating_operators}
\frac{d}{dt}\tilde{a}_{\sigma}(t)=R(t)\left[i\hat{\Lambda},\,\hat{a}_{\sigma}\right]R^{\dagger}(t)=-i\tilde{\Omega}_{\sigma}\,\tilde{a}_{\sigma}\rightarrow \tilde{a}_{\sigma}(t)=e^{-i\tilde{\Omega}_{\sigma}t}\tilde{a}_{\sigma}(0)=e^{-i\tilde{\Omega}_{\sigma}t}\hat{a}_{\sigma},
\end{equation}
one obtains that the rotated density operator evolves through the following master equation
\begin{equation}\label{eq:rotated_master_equation}
\frac{d}{dt}\tilde{\rho}=-i[\hat{\mathcal{H}}_{tot}(t)-\hat{\Lambda},\,\tilde{\rho}] + \sum_{\sigma}\Gamma_{\sigma}\mathbb{D}[\hat{a}_{\sigma};\,\tilde{\rho}],
\end{equation}
where
\begin{equation}
\begin{split}
\hat{\mathcal{H}}_{tot}(t)-\hat{\Lambda}&=(\tilde{\omega}_p-\tilde{\Omega}_p)\hat{a}^{\dagger}_{p}\hat{a}_p+(\tilde{\omega}_s-\tilde{\Omega}_s)\hat{a}^{\dagger}_{s}\hat{a}_s+(\tilde{\omega}_i-\tilde{\Omega}_i)\hat{a}^{\dagger}_{i}\hat{a}_i+\\
&+i\sqrt{\gamma}(\,F_p\,\hat{a}_{p}^{\dagger}-F^*_p\,\hat{a}_p)+ i\sqrt{\gamma}(\,F_s\,\hat{a}_{s}^{\dagger}e^{i(\tilde{\Omega}_s-\,\Omega_s)\,t}-F^*_s\,\hat{a}_s e^{-i(\tilde{\Omega}_s\,-\Omega_s)\,t})+\\ 
&+\sum_{k,\,l,\,m,\,n=p,\,s,\,i} \left(\hbar g_{klmn}\hat{a}_{k}^{\dagger}\hat{a}_{l}^{\dagger}\hat{a}_{m}\hat{a}_{n}e^{i(\tilde{\Omega}_k+\tilde{\Omega}_l-\tilde{\Omega}_m-\tilde{\Omega}_n)t}+h.c.\right).
\end{split}
\end{equation}
Let us now consider the rotating frame defined by the following constraints:
\begin{equation}
\tilde{\Omega}_s=\Omega_s \, ,
\end{equation} 
and
\begin{equation}
\tilde{\Omega}_i=2\tilde{\Omega}_s-\tilde{\Omega}_p \, ,
\end{equation}
the former corresponding to choose a reference frame rotating at $\Omega_s$ for the $s$-resonance, and the latter imposing energy conservation for the nonlinear terms converting two seed photons into a pump-idler pair, i.e., the processes considered in the main text. In this frame, by making a standard rotating-wave approximation (which corresponds to neglecting any off-resonant interaction term), the previous Hamiltonian model reduces to 
\begin{equation}
\begin{split}
\hat{\mathcal{H}}_{tot}(t)-\hat{\Lambda}&\approx(\tilde{\omega}_p-\tilde{\Omega}_p)\hat{a}^{\dagger}_{p}\hat{a}_p+(\tilde{\omega}_s-\Omega_s)\hat{a}^{\dagger}_{s}\hat{a}_s+(\tilde{\omega}_i+\tilde{\Omega}_p-2\Omega_s)\hat{a}^{\dagger}_{i}\hat{a}_i+\\
&+i\sqrt{\gamma}(\,F_p\,\hat{a}_{p}^{\dagger}-F^*_p\,\hat{a}_p)+ i\sqrt{\gamma}(\,F_s\,\hat{a}_{s}^{\dagger}-F^*_s\,\hat{a}_s )+\\ 
&+\sum_{\sigma=p,\,s,\,i}\hbar g_{\sigma\sigma\sigma\sigma} \hat{a}^{\dagger\,2}_{\sigma}\hat{a}^{2}_{\sigma}+ \sum_{k\neq l;k,l=p,\,s,\,i} \hbar g_{klkl}\hat{a}^{\dagger}_{k}\hat{a}_k \hat{a}^{\dagger}_{l}\hat{a}_l+ \hbar g_{nl}(\hat{a}^{\dagger\,2}_{s}\hat{a}_{i}\hat{a}_p +h.c)+\hat{V}(t)\\
\end{split}
\end{equation}
where we set $g_{ssip}\equiv g_{nl}$ for consistency with the notation used in the main text. \\
Evidently, the model still depends on the $p$-resonance. In order to get rid of any dynamical dependence on the $p$-operators, we apply the following argument. According to the discussion made in the previous section, the steady-state reached by the system in the continuous wave regime is independent of the particular protocol used for driving the system, i.e., protocol ``A" or ``B". Nevertheless, since we are looking for the stationary regime defined by  
\begin{equation}
\vert F_p\vert^2 \gg \vert F_s \vert^2,
\end{equation}
we do believe that, in order to build some intuition about the steady-state properties, it is convenient to consider a protocol of type ``A". If this is the case, at $t=t_1$ the system is in vacuum, and by switching the $F_p$ field on, it will be driven towards the steady-state of the master equation defined in Eq. \ref{eq:rotated_master_equation} for $F_s=0$. We argue that, due to the structure of the Hamiltonian, such a state will be a product state with the $p$-resonance somehow populated, and with the seed-idler subsystem empty. Indeed, no term in the Hamiltonian (due to the shifts between levels) is capable of transfering population from the $p$-resonance to the others. In other words, with reference to Fig.~\ref{fig:protocol_A}, for $t\approx t_c\gg t_1$ the system density operator will be $\tilde{\rho}(t)=\rho_{p;\,ss}\otimes \rho^{vacuum}_{s,i}$, with $\rho^{vacuum}_{s,i}$ denoting the vacuum state density operator of the seed-idler subsystem. In this scenario, we have that $\rho_{p,\,ss}$ is the solution of
\begin{equation}\label{eq:steady_state_kerr_model}
\tilde{\mathcal{L}}[\rho_{p;\,ss}]=-i[\hbar(\tilde{\omega}_p-\Omega_p) \hat{a}_{p}^{\dagger}\hat{a}_p+\hbar g_{pppp}\hat{a}_{p}^{\dagger\,2}\hat{a}_{p}^{2}+i\sqrt{\gamma}(\,F_p\,\hat{a}_{p}^{\dagger}-F^*_p\,\hat{a}_p),\,\rho_{p;\,ss}]+\Gamma_{p}\mathbb{D}[\hat{a}_{p};\,\rho_{p;\,ss}]=0 \, .
\end{equation}
Hence, by switching the second driving on, since we have that $\vert F_s\vert^2 \ll\vert F_p \vert^2$, it is reasonable to assume that the $p$-resonance will only be slightly affected by the photons injected in the seed-idler subsystem. As a consequence, the steady-state approached in the continuous-wave regime for $t\gg t_c$ can be approximated by a density operator that is still in factorized form. More precisely, we do expect the system to approach a configuration $\tilde{\rho}_{ss}\approx \rho_{p;\,ss}\otimes \rho_{s,i;\,ss}$, with $\rho_{s,i;\,ss}$ corresponding to the steady-state density operator that is solution of the following master equation:
\begin{equation}\label{eq:effective_master_equation_seed_idler}
\frac{d}{dt}\rho_{s,\,i}=-i\left[\hat{\mathcal{H}}_{s,\,i},\,\rho_{s,\,i}\right]+ \Gamma_{s}\mathbb{D}[\hat{a}_{s};\,\rho_{s,\,i}]+\Gamma_{i}\mathbb{D}[\hat{a}_{i};\,\rho_{s,\,i}],
\end{equation}  
with 
\begin{equation}\label{eq:effective_hamiltonian}
\begin{split}
\hat{\mathcal{H}}_{s,\,i}=&(\tilde{\omega}_s-\Omega_s+\delta_{XPM,\,p,\,s})\hat{a}^{\dagger}_{s}\hat{a}_s+(\tilde{\omega}_i+\Omega_p-2\Omega_s+\delta_{XPM,\,p,\,i})\hat{a}^{\dagger}_{i}\hat{a}_i+\\
&+i\sqrt{\gamma}(\,F_s\,\hat{a}_{s}^{\dagger}-F^*_s\,\hat{a}_s )+\hbar g(\alpha_{p}\hat{a}^{\dagger\,2}_{s}\hat{a}_{i} +\eta^*_{p}\hat{a}^{\dagger}_{i}\hat{a}^{2}_{s})+\\ 
&+\sum_{\sigma=\,s,\,i}\hbar g_{\sigma\sigma\sigma\sigma} \hat{a}^{\dagger\,2}_{\sigma}\hat{a}^{2}_{\sigma}+\hbar g_{sisi}\hat{a}^{\dagger}_{s}\hat{a}_s \hat{a}^{\dagger}_{i}\hat{a}_i,\\
\end{split}
\end{equation}
where $\delta_{XPM,\,p,\,\sigma}=g_{p\sigma p\sigma}\langle\hat{a}^{\dagger}_{p}\hat{a}_p\rangle_{p,\,ss}$ denotes the frequency shift related to the cross-phase modulation effects induced by the $p$-resonance on the $\sigma$-resonance and with $\alpha_{p}=\langle \hat{a}_p \rangle_{p,\,ss}$ denoting the field amplitude in the $p$-resonance. Before going any further, it is worth spending some words on the effective parameters entering the Hamiltonian model. On the one hand, by construction, $\langle\hat{a}^{\dagger}_{p}\hat{a}_p\rangle_{p,\,ss}$ is a real quantity and there is no ambiguity in using its value in the Hamiltonian; on the one other hand, the field amplitude $\alpha_{p}$ is, in general, complex, i.e. $\alpha_{p}=\vert \alpha_{p}\vert e^{i\phi}$, with $\vert \alpha_{p}\vert$ and $e^{i\phi}$ denoting the absolute value and the phase of $\alpha_{p}$. Such a phase is irrelevant for the dynamics. Indeed, such a phase can always be removed from the master equation by simply redefining the $\hat{a}_i$ operator, i.e., by using $\hat{a}_i\to e^{i\phi}\hat{a}_i$.\\
In particular, in the previous expressions, $\langle \hat{O}\rangle_{p,\,ss}\equiv \mbox{Tr}[\hat{O}\rho_{p;\,ss}]$ denotes the expectation value of the operator $\hat{O}$ on the state $\rho_{p;\,ss}$, that, interestingly, have some closed expression in terms of hypergeometric functions for the Lindblad evolution in Eq. \ref{eq:steady_state_kerr_model}, see e.g. \cite{Drummond_1980,Meaney2014,PhysRevA.94.033841}.\\ 
If we set $\omega_{\sigma}=\tilde{\omega}_{\sigma}+\delta_{XPM,\,p,\,\sigma}$, and suppose, without loss of generality, that
\begin{equation}
\omega_i+\omega_p=2\omega_s,
\end{equation} 
we have that the Hamiltonian model in Eq.\ref{eq:effective_hamiltonian}, up to corrections accounting for self-phase modulation and cross-phase modulation effects between the seed and ilder resonances, reduces to the one reported in the main text. Notice, however, that since such corrections are proportional to the bare coupling constant $g_{ssss}$, $g_{iiii}$ and $g_{sisi}$ which are by nature orders of magnitude smaller than the effective coupling $g_{nl}$, they can be safely neglected. 

\section{Steady-state observables for the seed-idler subsystem}\label{app:E}
In this section we report some details about the numerical methods used to determine the steady-state configuration and show some numerical results describing the behavior of the steady-state mean occupation of the seed resonance for values of the parameters not displayed in the main text.\\

\subsection{Convergence to the steady-state expectation values: finite-$N$ simulations}
\begin{figure}
    \centering
    \subfigure[$F_s/\sqrt{\Gamma_s}=1.0$ ]{\includegraphics[scale=0.45]{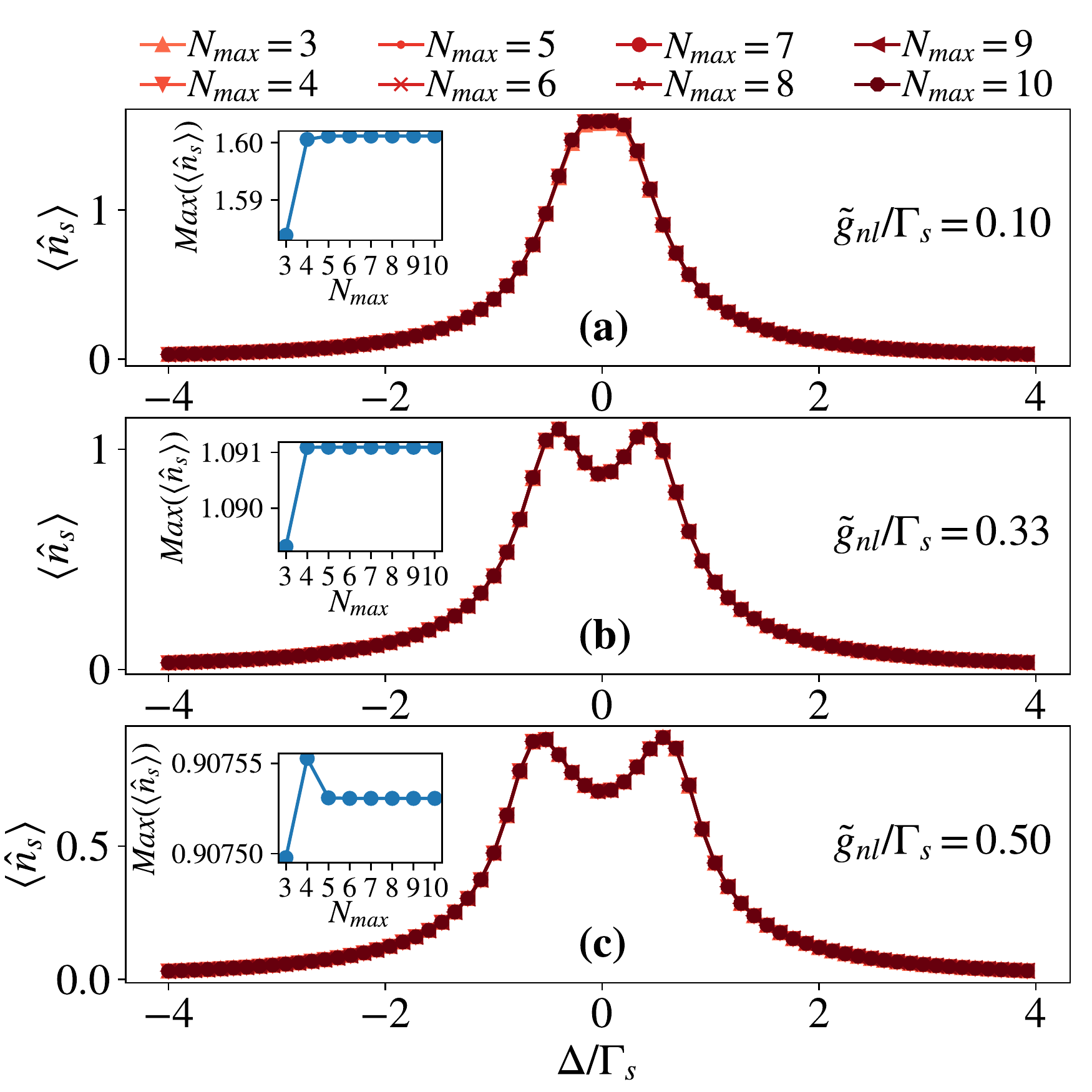}\label{subfig:F1.0}}
    \subfigure[$F_s/\sqrt{\Gamma_s}=2.5$ ]{\includegraphics[scale=0.45]{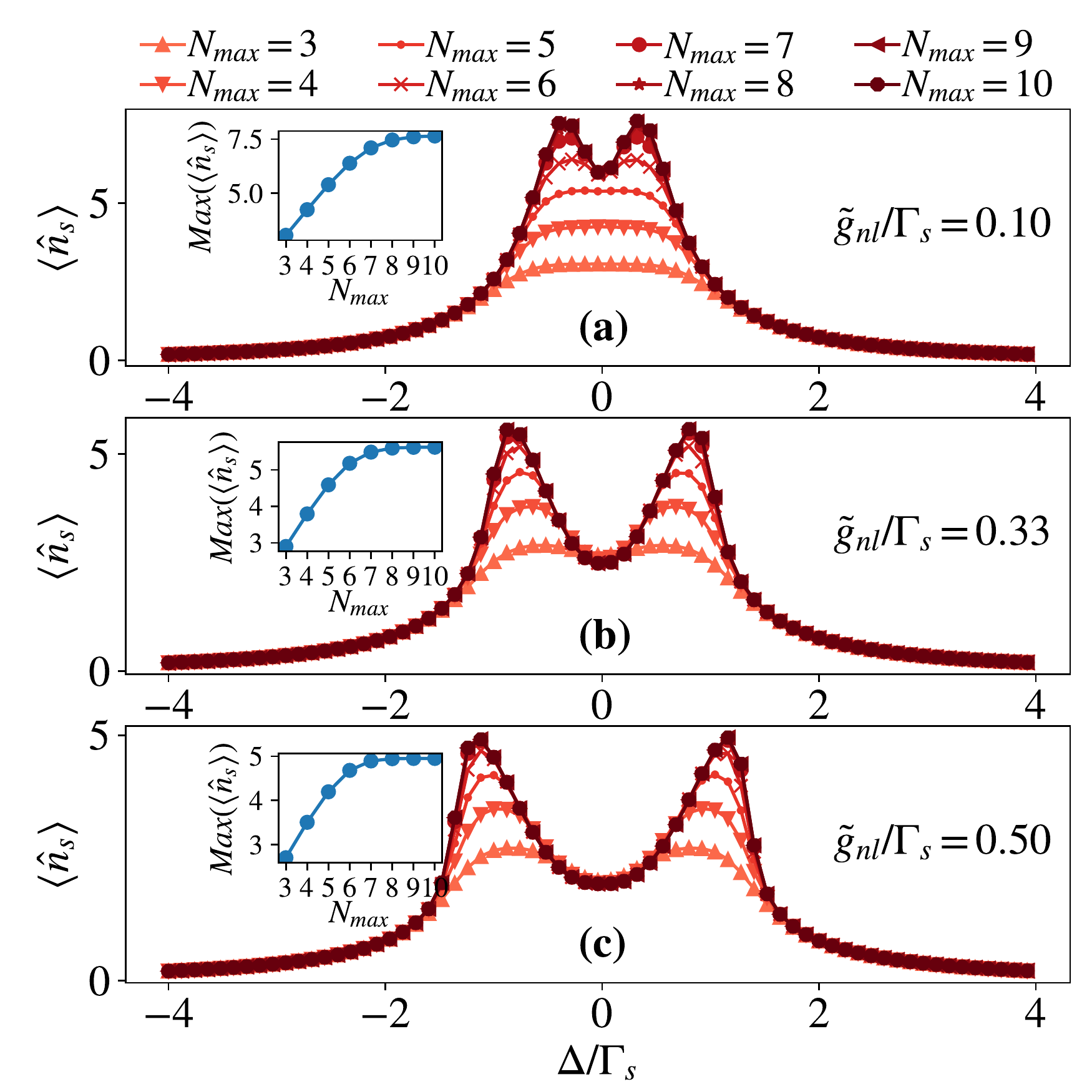}\label{subfig:F2.5}}
    \caption{Behavior of the steady-state mean occupation of the seed resonance $\langle \hat{n}_s\rangle$ as a function of the detuning $\Delta/\Gamma_s$, for different values of the effective nonlinear interaction $\tilde{g}_{nl}/\sqrt{\Gamma}$ (specified in each panel) and for $F_s/\sqrt{\Gamma_s}=1.0$ (\ref{subfig:F1.0}) and for $F_s/\sqrt{\Gamma_s}=2.5$ (\ref{subfig:F2.5}). The different curves in the main panels describe results obtained for different values of $N_{max}$ reported in the top legend. In each panel, we report also the behavior of $Max(\langle \hat{n}_s\rangle)$ as a function of $N_{max}$.}
    \label{fig:scaling_N}
\end{figure}

In the absence of losses and, in general, in the absence of any decay mechanism, boson systems in second quantization are described by the infinite dimensional space spanned by Fock states, i.e. number states. In the presence of losses, since high occupation states decay in favour of lower number states, one has that density operators can be approximate by means of a finite number of Fock states. To the best of our knowledge, such a number cannot be determined \emph{a priori}, and in order to characterize the system properties in the stationary regime one needs to look at the scaling of observables while increasing the number of states used in the representation of the Lindblad operator. Results shown in the main text and in this supplementary file have been obtained by adopting the following prescription: we truncate the Fock basis of the idler resonance at $N_{max}$ excitations, and the seed one at $2N_{max}$ excitations, with $N_{max}$ being a positive integer. The motivation for choosing a different truncation on the two resonances is related to the structure of the Hamiltonian. Indeed, since we considered only the effects determined by a direct driving on the $\omega_s$ resonance and since population is transferred to the idler resonance only via the nonlinear interaction (which converts two seed photons into an idler photon), in stationary conditions an idler state with $m$ photons is occupied if and only if a seed state with $2m$ photons is populated too. In particular, by proceeding this way one has that the number of states used to represent density operators of the seed-idler subsystem is given by $M(N_{max})\equiv(2N_{max}+1)(N_{max}+1)$, where the ``+1" comes from counting also the vacuum state. For any given $N_{max}$ and any point in the parameter space, we then determined, via standard linear algebra algorithm, the unique configuration annihilated by the Lindblad superoperator, that is the fixed point of Eq.\ref{eq:effective_master_equation_seed_idler} (that is by solving the linear system $\mathcal{L}[\rho_{ss}]=0$). \\
An example of scaling observed while increasing $N_{max}$ is reported in Fig. \ref{fig:scaling_N}. There, we show some numerical results obtained for the steady-state mean occupation of the seed resonance, that is $\langle \hat{n}_s\rangle$, as a function of the detuning $\Delta/\Gamma_s$, for different values of the effective nonlinear interaction $\tilde{g}_{nl}/\sqrt{\Gamma}$ (specified in each panel). In particular, results in Fig \ref{subfig:F1.0} and Fig \ref{subfig:F2.5} describe the convergence to the steady-state expectation value for $F_s/\sqrt{\Gamma_s}=1.0$ and $F_s/\sqrt{\Gamma_s}=2.5$. As suggested by the inset plots in the panels describing the behavior of $Max(\langle \hat{n}_s\rangle)$, due to the weaker driving rate, data for $F_s/\sqrt{\Gamma_s}=1.0$ can be considered at convergence even at $N_{max}\sim 5$. On the contrary, for $F_s/\sqrt{\Gamma_s}=2.5$, data can be considered at convergence for $N_{max}\sim 8-9$.\\ 

\subsection{Behavior of the seed mean occupation at increasing power}
\begin{figure}
    \centering
    \includegraphics[scale=0.6]{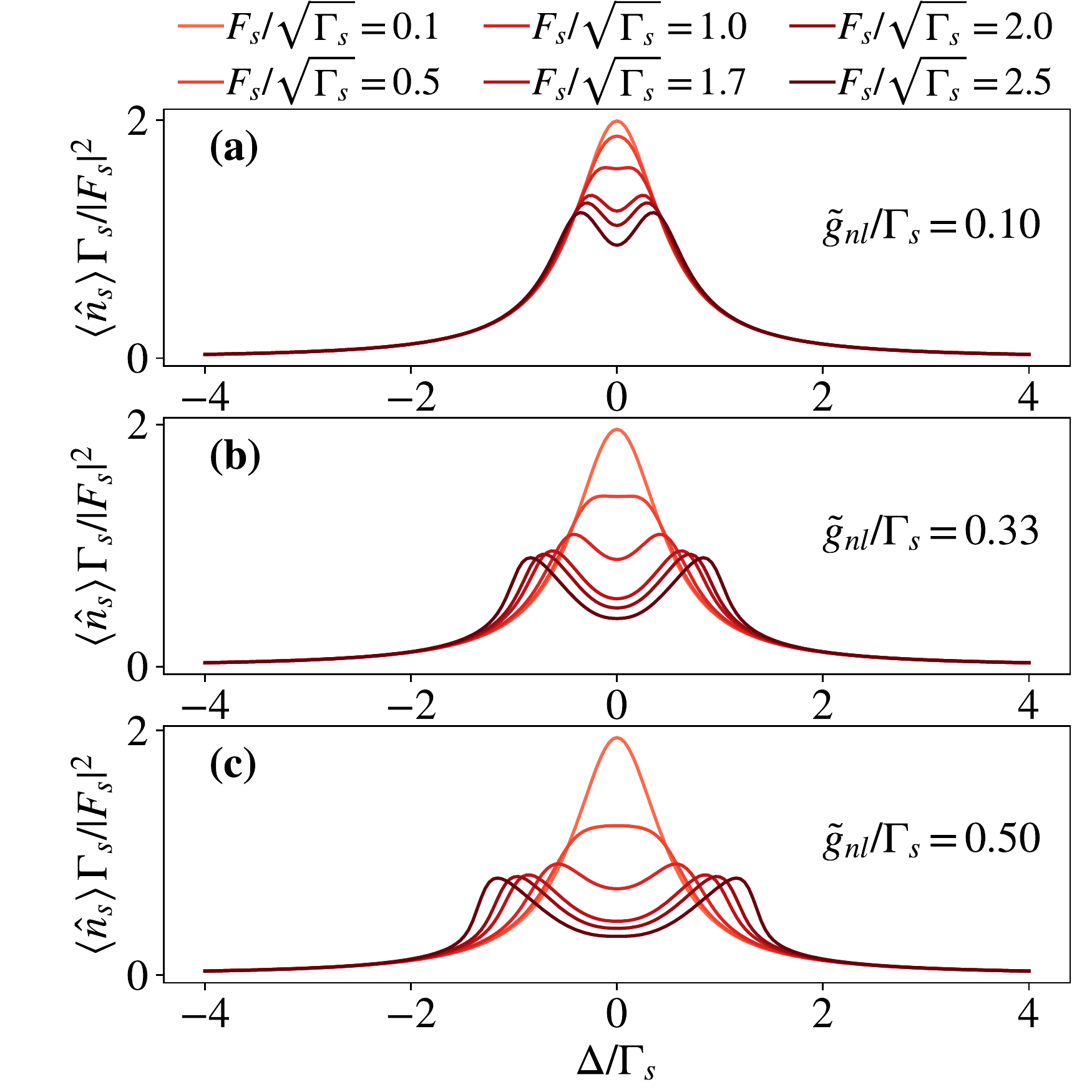}
    \caption{Behavior of the ratio between the steady-state mean occupation of the seed resonance, that is $\langle\hat{n}_s\rangle$, and the normalized driving intensity $\vert F_s\vert^2/\Gamma_s $, as a function of the detuning  $\Delta/\Gamma_s$. The different panels correspond to different values of the effective effective nonlinearity $\tilde{g}_{nl}/\Gamma_s$.}
    \label{fig:splitting_lines}
\end{figure}

In this section we show some numerical results describing the behavior at increasing $F_s$ of the mean occupation of the seed resonances. In particular, in Fig. \ref{fig:splitting_lines} we show the behavior of such quantity (data displayed correspond to  $N_{max}=10$) divided by $\vert F_s \vert^2/\Gamma_s$ (which in general provides an estimate of the mean number of photons stored within a linear driven-dissipative resonator). As it is possible to see, on the increasing of the input power acting on the seed resonance ($\propto \vert F_s \vert^2 $), the system displays a pronounced line-splitting. At low driving, the mean-steady state occupation is characterized by a single Lorentzian-peak, located at $\Delta=0$, i.e. $\Omega_s=\omega_s$. On the increasing of the driving intensity of CW at $\Omega_s$, in agreement with results previously appeared in literature, see e.g. \cite{Ramelow2019}, two peaks appear in the seed response, which is a signature of the presence of a strong nonlinear interaction acting in the seed-idler subsystem. 

\end{document}